\documentstyle[12pt]{article}

\def\beqa{\begin{eqnarray}}
\def\eeqa{\end{eqnarray}}
\def\beq{\begin{equation}}
\def\eeq{\end{equation}}

\begin{document}

\begin{titlepage}
        \title{Higher order corrections to lensing parameters
for extended gravitational lenses}
\author{Salvatore Capozziello\thanks{capozziello@sa.infn.it },~~
Virginia Re\thanks{re@sa.infn.it} \\ {\em Dipartimento di  Fisica
"E.R. Caianiello"} \\
 {\em Universit\'a di Salerno, 84081 Baronissi (Sa), Italy.} \\
 {\em Istituto Nazionale di Fisica Nucleare, Sez. di Napoli, Italy.} \\ }
\date{\today}
\maketitle

\begin{abstract}
We discuss the contribution  to
the characteristic lensing quantities, i.e.
the deflection angle and Einstein radius, due to
 the higher order terms (e.g. the gravitomagnetic terms)
considered in the lens potential.
 The cases we analyze are the
singular isothermal sphere and the disk of spiral galaxies. It is
possible to see
that  the perturbative effects could be of the order $10^{-3}$
with respect to the ordinary terms of weak field and thin lens approximations,
 so that it is not a far hypothesis
to obtain evidences of them  in a next future by suitable experiments.

\end{abstract}
 \vspace{10.mm}
 PACS number(s): 04.50.+h, 04.20. Cv, 98.80. Hw \\
 \vspace{5.mm}
 Keyword(s):  gravitational lensing,  galaxy models.
 \vfill
 \end{titlepage}

\section{\normalsize\bf Introduction}

In the last decades, gravitational lensing has revealed a powerful
tool in astrophysics and cosmology so that a newborn branch of
astronomy has been named "gravitational astronomy". Some basic
assumptions are always taken into account in the theory: the first
is the weak field approximation. It is due to the fact that the
most of  gravitational systems considered, as lenses and sources,
are well beyond their Schwarzschild radii so that strong regimes
can be ignored. Besides, thin lens approximation is often used
since the thickness of deflecting systems is very small  with
respect to the distances between observer, lens, and source
respectively. A further simplification is related to the fact that
deflection angles are, in general, small. The result of these
approximations is that the theory of gravitational lensing is
quite simple \cite{ehlers} and the main interests are devoted to
the development of applications.

However, gravitational lensing must not be conceived as a weak
field phenomenon since the bending and the looping of light rays
are predictions of  "full''  General Relativity theory. Besides
this fact of principle, several facts and observations ask for the
enlargement to any order of approximation. For example, when the
lenses are compact objects, as black holes or pulsars, the above
approximations never hold so that more general and involved
schemes have to be adopted. In these cases, the gravitational
systems are close to their Schwarzschild radii and the weak field
approximation fails \cite{ellis},\cite{bozza}. A full treatment of
lensing phenomena gives rise to a transcendent lens equation which
has to be handled with some care in order to be solved
\cite{bozza}.

Further considerations deserve the weak field limit which could
give interesting results if it is not simply considered at lowest
orders. In \cite{capozziello}, it was shown that taking into
account corrections of order $v/c$ gives rise to nonnegligible
effects which becomes more and more interesting in a cosmological
context. Manzano and Montemayor \cite{manzano} develops the full
theory of the propagation of light in a gravitational background
showing that several effects, as polarization, depends on  the
intrinsic feature of the gravitational field of the lens. It is
worthwhile to note the so called Rytov effect \cite{rytov}, i.e.
the rotation of the plane of polarization of the incident beam
which is generated by the  gravitational field of a rotating mass.
In this effect, the angular momentum gives contribution to the
deflection angle. In general, as it is widely discussed in
\cite{capozziello}, taking into account the gravitomagmetic term
introduces corrections which are useful if we consider the proper
motions of the lens. Their magnitude can be relevant depending on
the ratio $v/c$.

As a further consideration, the study of higher order perturbative
terms is the link between weak and strong regimes which can allow
the  full parametrization of lensing theory.

Finally, a close comparison between General Relativity and other
relativistic theories of gravity can be led on the base of higher
order effects since  the differences between them could not emerge
in the lowest order of the weak field limit
\cite{will},\cite{sirousse},\cite{calchi}.

In this paper, we take into account the gravitomagnetic
contributions to the lensing quantities, as the deflection angle
and the Einstein radius, for extended gravitational lenses of
astrophysical interest as the singular isothermal sphere and the
disk of spiral galaxies. In both cases, the proper motions (i.e.
the velocity dispersion and the circular velocity) have to be
taken into account and, in some peculiar cases (e.g. AGN, or other
kinds of extreme active galaxies) their contributions could be far
to be trivially ignored.

However, the singular isothermal sphere as lens was considered by
many authors. For example, Gurevich and Zybin \cite{gurevich} use
it to study structures of cold dark matter at small scale regions.
They give a constraint for the minimal dark matter objects (which
should be in the range $M_X\sim 10^{-4}\div 10^{-2} M_{\odot}$)
and another for the particle candidate (neutralino) $m_X\sim 10\pm
7$ GeV.  Sazhin et al. \cite{sazhin} adopt singular isothermal
sphere models to study microlensing by non-compact invisible
objects. They discuss the OGLE collaboration results towards Large
Magellanic Cloud which point out the presence of dark bodies with
a mass of the order of $0.1 M_{\odot}$. The authors suggest,
besides the standard hypothesis that such objects are brown
dwarfs, other hypotheses which includes mirror stars, black holes,
and exotic stars which consist of cold dark matter particles. In
particular, they discuss the last possibility which is attractive
from the point of view of modern particle physics and find that
some observational data can be consistent with it.

In Sect.2, we discuss the case of singular isothermal sphere and
in Sect.3, the disk of spiral galaxies is taken into account. The
 results are discussed in Sect.4.

\section{\normalsize\bf The singular isothermal sphere}
Isothermal sphere is the simplest  model used to describe the mass
function of the haloes of galaxies and to derive the potential of
elliptical galaxies \cite{binney}. As a lens model, it is the
further step after the pointlike Schwarzschild lens. The internal
motions (e.g. the velocity dispersions, proper motions of the
stars, etc.) give rise to nontrivial effects capable of supporting
the dynamics of the real systems \cite{binney}. If the system
described by an isothermal sphere acts as a lens,
 these effects can lead to  gravitomagnetic corrections
 which could be quantitatively significant.

 The mass density for a singular isothermal sphere
is given by

\begin{equation}\label{1}
  \rho=\frac{\sigma_{v}^{2}}{2 \pi G x^{2}}
\end{equation}

with
\begin{equation}\label{2}
  \vec{x}=\vec{\xi}+l \vec{e}_{in}
\end{equation}
where $\sigma_{v}$ is the velocity dispersion of the lens, $\xi$
is the distance from the centre of the sphere, $\vec{e}_{in}$ is a
unitary vector in the initial direction of light. Gravitational
potential is given by

\begin{equation}\label{3}
  \phi(x)=-G \int d^{3}x'\frac{\rho(x)}{|\vec{x}-\vec{x'}|}=
  -\frac{\sigma_{v}^{2}}{2 \pi}\int
  d^{3}x'\frac{1}{x'^{2}}\frac{1}{|\vec{x}-\vec{x'}|};,
\end{equation}
which is

\begin{equation}\label{4}
  \phi(x)=-2 \sigma_{v}^{2}\ln\left(\frac{x}{R}\right)=
-2\sigma_{v}^{2}\ln\left(\frac{\xi+le_{in}}{R}\right)
\end{equation}
where $R$ is a cut-off distance introduced to eliminate  the
singularity in the origin.

A vector potential can be defined as
\begin{equation}\label{5}
 \vec{V}=\vec{v}\phi
\end{equation}
where $\vec{v}$ is a velocity.  The k-component  results

\begin{equation}\label{6}
  V_{k}=-2 \sigma_{v}^{2} v_{k}\ln\left(\frac{\xi+l
  e_{in}}{R}\right)\,.
\end{equation}

The deflection angle of light, taking into account also the
potential vector term \cite{ehlers},\cite{capozziello} is given by
\begin{equation}\label{7}
 \vec{\alpha}=\frac{2}{c^{2}}\int
 \vec{\nabla}_{\bot}\phi dl -\frac{4}{c^{3}}\int
 (\vec{e}\wedge(\vec{\nabla}\wedge\vec{V}))dl\,,
\end{equation}
where $dl$ is the Euclidean line element. The last term in the
right-hand side is the gravitomagnetic correction. If we solve for
every single vector component, evaluating the integrals between
$0$ and $\infty$, we get

\begin{equation}\label{8}
  \alpha_{k}=-\frac{4\sigma_{v}^{2}\xi_{k}}{c^{2}}\int_{0}^{\infty}\frac{1}{\xi^{2}+l^{2}}
dl-\frac{8\sigma_{v}^{2}
v_{k}}{c^{3}}\int_{0}^{\infty}\frac{l}{\xi^{2}+l^{2}} dl
\end{equation}

which results
\begin{equation}\label{9}
 \alpha_{k}=-\frac{2\sigma_{v}^{2}\pi}{c^{2}}\frac{\xi_{k}}{\xi}-\frac{4\sigma_{v}^{2}
v_{k}}{c^{3}}\ln\left[1+\left(\frac{R}{\xi}\right)^{2}\right]\,.
\end{equation}

The last term in (\ref{9})is  due to the gravitomagnetic
correction and it clearly depends by the ratios  $v_{k}/{c}$ (i.e.
the kinematics) and $R/\xi$ (i.e. the geometry). It is
straightforward to see that the correction is significant only for
appreciable values of these ratios. For example, it is easy to
estimate that  for $v_{k}/c\simeq 10^{-(2\div 3)}$ and
$R\sim\sqrt{2}\xi$ we could appreciate some effects. This means
that high proper motions and the impact parameters of the light
beams comparable to the physical sizes of the lenses can give rise
to appreciable gravitomagnetic corrections.

\section{\normalsize\bf The disk of spiral galaxies}
Gravitomagnetic corrections to the disk potential of a spiral
galaxy are another interesting case. The k-component of the
deflection angle is

\begin{equation}\label{11}
\alpha_{k}^{grav}(\xi)=-\frac{4}{c^{3}}\int[\vec{e}\wedge(\vec{\nabla}\wedge\vec{V})]_{k}dl=
-\frac{4}{c^{3}}\int\left[\partial_{k}(\vec{e}\cdot\vec{V})-(\vec{e}\cdot\vec{\nabla})_{k}\cdot\vec{V}\right]dl.
\end{equation}
 Solving the integral in polar coordinates
\begin{equation}\label{12}
dl\equiv(dx,dy)\rightarrow(d\xi,\xi d\theta)
\end{equation}
and taking in account that the potential does not depend on
$\theta$ for symmetry properties, the integration becomes simply
depending by $\xi$. It is easy to find that

\begin{equation}\label{13}
  \alpha_{k}^{grav}(\xi)=\frac{4}{c^{3}}v_{k}\int\frac{d\psi(\xi)}{d\xi}d\xi
\end{equation}
in which $\psi(\xi)$ is the gravitational potential. The mass
distribution of a spiral galaxy is
\begin{equation}\label{14}
  M(\xi)=2\pi
  \xi_{c}^{2}\Sigma_{0}\left[1-\left(1+\frac{\xi}{\xi_{c}}\right)e^{-\frac{\xi}{\xi_{c}}}\right]
\end{equation}
where $\xi_c$ is a scale length normally projected along the line
of sight \cite{ehlers}.

The deflection angle for a disk-like deflector is:

\begin{equation}\label{15}
 \alpha(\xi)=\frac{4G M(\xi)}{c^{2}\xi}=
 \frac{8 \pi G \xi_{c}^{2}\Sigma_{0}}{c^{2}}
 \left[\frac{1}{\xi}-\frac{\left(1+\frac{\xi}{\xi_{c}}\right)
 e^{-\frac{\xi}{\xi_{c}}}}{\xi}\right]\;.
\end{equation}
The potential is given by

\begin{equation}\label{16}
  \psi(\xi)=-\left(\frac{D_{ds}}{D_{d}D_{s}}\right)\int\alpha(\xi)d\xi
  +\psi_{0}\,,
\end{equation}
and then
\begin{equation}\label{17}
  \psi(\xi)=\psi_{0}-\left(\frac{D_{ds}}{D_{d}D_{s}}\right)\frac{8\pi G \xi_{c}^{2}\Sigma_{0}}{c^{2}}
  \left\{e^{-\frac{\xi}{\xi_{c}}}+\frac{\xi}{\xi_{c}}-\frac{1}{4}\left(\frac{\xi}{\xi_{c}}\right)^{2}
  +\frac{1}{18}\left(\frac{\xi}{\xi_{c}}\right)^{3}+...\right\}
\end{equation}
where $D_{ds},D_{d},D_{s}$ are the deflector-source, the
deflector-observer, and the source-observer distances
respectively.

 Immediately, we see that the correction to $\alpha$ due to
the gravitomagnetic term is given by

\begin{equation}\label{18}
  \alpha^{grav}_{k}(\xi)=\frac{4}{c^{3}}v_{k}\psi_{0}-\left(\frac{D_{ds}}{D_{d}D_{s}}\right)\frac{32 v_{k} \pi G \xi_{c}^{2}\Sigma_{0}}{c^{5}}
  \left\{e^{-\frac{\xi}{\xi_{c}}}+\frac{\xi}{\xi_{c}}-\frac{1}{4}\left(\frac{\xi}{\xi_{c}}\right)^{2}
  +\frac{1}{18}\left(\frac{\xi}{\xi_{c}}\right)^{3}+...\right\}
\end{equation}
which is, in a compact form,
\begin{equation}\label{19}
\alpha^{grav}_{k}(\xi)=\frac{4}{c^{3}}v_{k}\psi_{0}-\left(\frac{D_{ds}}{D_{d}D_{s}}\right)
\frac{32v_{k} \pi G \xi_{c}^{2}\Sigma_{0}}{c^{5}}
\left[{e^{-\frac{\xi}{\xi_{c}}}-\sum_{n=1}^{\infty}(-1)^{n}\frac{1}{nn!}
\left(\frac{\xi}{\xi_{c}}\right)^{n}}\right]\,.
\end{equation}
Also in this case, the role of the ratios ${\displaystyle
\frac{v_k}{c}}$ and ${\displaystyle \frac{\xi}{\xi_c}}$ is leading
to appreciate the correction.

\section{\normalsize\bf Discussion and Conclusions}

In this paper, we have analyzed, in the framework of gravitational
lensing theory,  the effects of  gravitomagnetic corrections on
the deflection angle in the case of extended mass lenses. In
\cite{capozziello}, gravitomagnetic corrections have been taken
into account for   point-like deflectors and,  in that case, the
deflection angle is corrected by a factor which  can be related to
the redshift of the lens. It is  clear that the possibility to
detect the correction depends on the fact that point-like lenses
should have a sufficiently high redshift (which could be intrinsic
in the case of high proper motions).

These results can be generalized to extended lens models  where
the effects of  proper motions can be very relevant and then the
vector potential terms in the perturbative expansion of
gravitational field are not negligible.

In particular, we have analyzed two mass models: the singular
isothermal sphere and the disk of spiral galaxies.

In the first case, we have found that the gravitomagnetic
correction depends on the ratios $v_{k}/{c}$ and $R/\xi$, that is
on the kinematics and the geometry of the system. Then it is still
possible to link this term to the redshift so that we can give a
quantitative evaluations  of gravitomagnetic corrections.  To give
relevant effects, the second term in Eq.(\ref{9}) should be of the
order $10^{-(2\div 3)}$ with respect to the first. These are the
limits set by the forthcoming  space and ground-based experiments
\cite{ulvestad}. It is easy to see that these constraints can be
achieved by taking into account exotic objects (e.g. AGN) as
lenses. It is worthwhile to stress the fact that such constraints
could be used also to study exotic non-compact invisible bodies
\cite{gurevich},\cite{sazhin} by taking into account their
kinematics.

 In the case of the
disk of spiral galaxies, we find that the  correction to the
deflection angle depends on exponential term (decreasing according
to a scale factor) and on  an infinite series of terms depending
on the same scale factor [see Eq.(\ref{18})].  Similarly to the
above situation, geometry and kinematics lead the corrections and
can be appreciated at certain scales. This fact is in agreement
with the results of several relativistic theories of gravity,
where the corrections to General Relativity are relevant depending
on the interaction scales (see e.g.
\cite{sirousse},\cite{calchi},\cite{mannheim},\cite{cap2}). In
those cases, the Newtonian potential has to be corrected by Yukawa
terms or power law series,  strictly dependent on some
characteristic scale (e.g. given by the mass of some interaction
boson). For example, considering the conformal gravity, Mannheim
et al. \cite{mannheim}  "derive'' the flat rotation curve of
spiral galaxies without using huge amounts of dark matter.
However, also lensing quantities are affected in these extended
theory of gravity (see \cite{sirousse},\cite{calchi}). For
example, in \cite{edery} a systematic analysis of light deflection
and time delay is performed in the context of Weyl gravity. The
main result is that all lensing quantities are corrected and
several  observations, which cannot be fitted in General
Relativity context, could be reinterpreted in this scheme. From
our point of view, higher order corrections in the series
expansion of gravitational field acquire a similar role since they
reproduce, at least from a formal point of view the same results.
In this sense, they  have to be considered if relativistic effects
are not negligible.

\vspace{2. mm}

\noindent {\bf Ackwledgement}\\ The authors wish to acknowledge
Fabrizio Tamburini for the useful suggestions which allowed to
improve the paper.


\begin{thebibliography}{99}

\bibitem{ehlers}  Schneider P., Ehlers J., Falco
E.E.,1992, Gravitational lenses. Springer-Verlag, Berlin.

\bibitem{ellis}  Virbhadra K.S., Ellis G.F.R., 2000, Phys. Rev. D {\bf 62},
084003.

\bibitem{bozza} Bozza V., Capozziello S., Iovane G., Scarpetta G., 2001,
Gen. Rel and Grav., {\bf 33}, 1.

\bibitem{capozziello}  Capozziello S., Lambiase G., Papini G.,
Scarpetta G., 1999, Phys. Lett. A, 254, 11.

\bibitem{manzano}  Manzano J. and  Montemayor R., 1997, Phys. Rev. D {\bf 57},
6378.

\bibitem{rytov}  Rytov S.M., 1938 Dokl. Akad. Nauk. SSSR {\bf 18},
263.
\bibitem{will} Will C.M., " Theory and Experiments in
gravitational Physics", Cambridge Univ. press, Cambridge, 1993.

\bibitem{sirousse} Sirousse-Zia H., 1998  Gen. Rel.and Grav. {\bf
30}, 1273.

\bibitem{calchi} Calchi Novati S., Capozziello S., and Lambiase
G., 2000  Grav. \& Cosm. {\bf 6}, 173.

\bibitem{gurevich} Gurevich A.V. and Zybin K.P., 1995 Phys. Lett.
A {\bf 208},  276.

\bibitem{sazhin} Sazhin M.V.,  Yagola A.G.,  and Yakubov A.V., 1996 Phys. Lett.
A {\bf 219},  199.

\bibitem{binney} Binney J., Tremaine S. "Galactic Dynamics'',
Princeton Univ. Press, Princeton 1987.

\bibitem{ulvestad}  Ulvestad J.S., 1999, ``Goals of the ARISE
Space VLBI Mission'', New Astronomy Reviews, Proceedings of the
4th EVN/JIVE Symposium.

\bibitem{mannheim} Mannheim P.D. and Kazanas D., 1989 Ap. Jou. {\bf
342}, 635. \\ Barabash O.V. and Shtanov Yu. V., 1999,  Phys. Rev.
D {\bf 60} 064008.

\bibitem{cap2} Capozziello S., De Martino S., De Siena S., Illuminati F.,
2001 Mod. Phys. Lett.  A 693.

\bibitem{edery} Edery A. and Paranjape M.B., 1998 Phys. Rev. D {\bf 58 }
024011.

\end{thebibliography}
\end{document}